\documentclass[conference]{IEEEtran}
 \pdfoutput=1 
\usepackage{cite}
\usepackage{amsmath,amssymb,amsfonts}
\usepackage{algorithmic}
\usepackage{graphicx}
\usepackage{tikz-layers}
\usepackage{textcomp}
\usepackage{xcolor}
\def\BibTeX{{\rm B\kern-.05em{\sc i\kern-.025em b}\kern-.08em
    T\kern-.1667em\lower.7ex\hbox{E}\kern-.125emX}}
\usepackage{tikz}
\tikzstyle{pinstyle} = [pin edge={to-,thin,black}]
\usetikzlibrary{calc}
\newlength\fheight
\newlength\fwidth
\usepackage{pgfplots}
\pgfplotsset{compat=1.17}
\usepackage{cite}
\usepackage{url}
\usepackage{amsmath,amsfonts,amssymb,amsthm, bm}
\usepackage{arydshln}
    \usepackage{subfig}

\usepackage{pifont}
\newcommand{\yes}{\text{\ding{51}}}
\newcommand{\no}{\text{\ding{55}}}
\usepackage{multirow}
\usepackage{rotating}
\newcommand{\ctp}[1]{\textit{#1}}

\DeclareMathSymbol{\smi}{\mathbin}{AMSa}{"39}
\usepackage{textcomp}
\usepackage[absolute]{textpos}

\newcommand{\copyrightstatement}{
    \begin{textblock}{15}(0.5,0.3)    
         \noindent
         \centering
         \textblockcolour{white}
         \footnotesize
         \copyright 2022 IEEE. Personal use of this material is permitted. Permission from IEEE must be obtained for all other uses, in any current or future media, including reprinting/republishing this material for advertising or promotional purposes, creating new collective works, for resale or redistribution to servers or lists, or reuse of any copyrighted component of this work in other works
    \end{textblock}
}

\begin{document}

\copyrightstatement
\title{Advanced Design Space Exploration for Joint Energy and Quality Optimization for VVC}

\author{\IEEEauthorblockN{Matthias Kr\"anzler, Andr\'e Kaup, and Christian Herglotz}
	\IEEEauthorblockA{Chair of Multimedia Communications and Signal Processing, \\
		Friedrich-Alexander-Universität Erlangen-N\"urnberg (FAU)\\
		\{matthias.kraenzler, andre.kaup, christian.herglotz\}@fau.de\\}}

\maketitle

\begin{abstract}

In recent studies, it could be shown that the energy demand of Versatile Video Coding (VVC) decoders can be twice as high as comparable High Efficiency Video Coding (HEVC) decoders. A significant part of this increase in complexity is attributed to the usage of new coding tools. By using a design space exploration algorithm, it was shown that the energy demand of VVC-coded sequences could be reduced if different coding tool profiles were used for the encoding process. This work extends the algorithm with several optimization strategies, methodological adjustments to optimize perceptual quality, and a new minimization criterion. As a result, we significantly improve the Pareto front, and the rate-distortion and energy efficiency of the state-of-the-art design space exploration. Therefore, we show an energy demand reduction of up to 47\% with less than 30\% additional bit rate, or a reduction of over 35\% with approximately 6\% additional bit rate.
\end{abstract}

\IEEEpeerreviewmaketitle

\section{Introduction}
\label{sec:intro}

Since the beginning of the internet, global IP traffic is constantly increasing. According to a report from 2019~\cite{CSI2019}, the total IP traffic equaled 122 petabytes per month in 2017 and was predicted to be 396 petabytes per month in 2022, if a compound annual growth rate of 26\% is assumed. By 2022, a share of 82\% is attributed to video content such as video-on-demand streaming or video conferencing. The growing demand for higher resolution content led to the development of the video standard Versatile Video Coding (VVC), which was standardized in 2020. With VVC it is possible to reduce the bit rate by 50\% in relation to its predecessor High-Efficiency Video Coding (HEVC) at an equal subjective quality~\cite{Bross2021}. This is achieved by using several new coding tools and an enhanced partitioning scheme. However, the increased compression efficiency is at the cost of increased complexity for both the encoder and the decoder. 

With Green Metadata~\cite{MPEG-Green} one can control the decoder complexity since it enables an additional communication channel from the decoder to the decoder. In the Green Metadata version for VVC, there are three methods specified that enable a request of the decoder for a complexity reduction. First, the reduction of decoder operations in the encoder. Secondly, spatial and temporal scaling. Finally, a coding tool configuration that can disable the usage of loop filters or the usage of intra prediction in B-frames. Furthermore, it is possible to request one of four user defined coding tool profiles (CTPs).

In~\cite{Kraenzler21}, it is shown that the energy demand of HEVC decoders is 40\% to 50\% lower than comparable VVC decoder implementations. Furthermore, the authors propose an algorithm that is based on a greedy strategy based design space exploration~(DSE) that derives energy efficient CTPs. This reduces the energy demand of VVC decoders by controlling the usage of specific coding tools in the encoder. Thereby, the authors show that the energy demand of VVC encoded bit streams is reduced by up to 50\% with an additional bit rate of up to 25\% in relation to the default usage of coding tools in VVC. As a result, the energy efficiency of the proposed CTPs was significantly improved over HEVC with a significantly lower additional bit rate. However, the algorithm was only optimized on minimal energy demand. In~\cite{Kraenzler2022}, the authors refined the proposed CTPs by selecting coding tools without changing the algorithm of~\cite{Kraenzler21}. It was shown that the additional bit rate is reduced and the energy reductions are still preserved for a similar encoding time as the \ctp{medium} preset of VVenC, which is an optimized VVC encoder implementation. However, there is no algorithmic solution to get CTPs that have a low bit rate increase.

In this paper, we provide an algorithmic and methodological solution to enhance the design space exploration. Therefore, we have the following contributions:
\begin{itemize}
\item Evaluation of the algorithm with optimized VVC software
\item Optimization on the perceptual quality by using VMAF
\item Enhanced optimization strategies for DSE
\item Proposal of multiple CTPs based on a joint optimization of energy and rate-distortion efficiency
\end{itemize}

The state-of-the-art DSE algorithm will be explained in Section~\ref{sec:Alg}. Afterward, we show the modifications and proposed optimization strategies in Section~\ref{sec:Modifications}. Then, we present the used setup and metrics in Section~\ref{sec:Setup}. In Section~\ref{sec:Evaluation}, we show the evaluated results and propose several coding tool profiles that optimize the energy efficiency of VVC and also jointly optimize the compression and energy efficiency. Finally, in Section~\ref{sec:conclusion} we will conclude this paper with a summary and an outlook for future research.

\section{State-of-the-art DSE Algorithm}
\label{sec:Alg}

For this paper, we use the state-of-art DSE algorithm of~\cite{Kraenzler21} as a baseline for our improvements. The method optimizes the decoder energy demand by switching the usage of selected tools in the encoder. Therefore, a coding tool profile (CTP) vector $\boldsymbol{u}$ is defined, which describes the usage characteristic of a set of coding tools. The goal is to find a CTP that has a minimal energy demand, which is expressed by the following optimization function
\begin{equation}
\label{eq:1}
\min_{\boldsymbol{u}}~\textrm{BDDE-PSNR}  = \min_{\boldsymbol{u}}~f\left(\boldsymbol{u} \right).
\end{equation}
 The reduction of the decoding energy demand is evaluated with the Bj{\o}ntegaard-Delta decoding energy (BDDE-PSNR) metric. This metric evaluates the energy savings at an equal PSNR and will be explained in detail in Section~\ref{sec:Setup}. The algorithm in~\cite{Kraenzler21} evaluates the energy efficiency of each coding tool first, and based on the energy efficiency in terms of BDDE-PSNR, changes the usage of each coding tool if the energy efficiency was improved. This is done in multiple iterations and for each iteration, the energy efficiency is evaluated in relation to the reference. Furthermore, for each coding tool that improves energy efficiency within one iteration compared to the iteration reference, the usage of the corresponding tool is changed for the reference CTP of the next iteration. As a consequence, the usage of single or multiple coding tools can be changed within one iteration. The algorithm terminates if the reference CTP of an iteration is the same as in a previous iteration.
 
 \section{Proposed Optimized Strategies for DSE}
\label{sec:Modifications}
 
In contrast to~\cite{Kraenzler21}, we optimize the energy demand on the metric Video Multimethod Assessment Fusion~(VMAF), which claims to predict the subjective quality with higher accuracy than PSNR~\cite{LiAaronKatsavounidisEtAl2016}. Therefore, we change the minimization criterion in this work to
\begin{equation}
\label{eq:2}
\min_{\boldsymbol{u}}~\textrm{BDDE-VMAF}  = \min_{\boldsymbol{u}}~f\left(\boldsymbol{u} \right),
\end{equation}
which improves the energy demand for the same perceptual quality in terms of VMAF.

According to the optimization strategy in~\cite{Kraenzler21}, it was possible to switch the usage of every coding tool that improved the energy efficiency. Therefore, we propose to limit the maximum number of allowed coding tools that can be changed after one iteration to one. Therefore, only the usage of the coding tool with the lowest BDDE value is changed, and the remaining coding tools are unchanged even though the energy efficiency of those coding tools is improved.

Furthermore, we improve the optimization criterion of~\eqref{eq:2} by jointly optimizing BDDE and BDR to derive CTPs that maximally improve the energy efficiency for a given additional bit rate. Therefore, we have the following optimization function for the joint optimization
\begin{equation}
\label{eq:3}
\min_{\boldsymbol{u}}~\left(\textrm{BDDE-VMAF}+\textrm{BDR-VMAF}\right)  = \min_{\boldsymbol{u}}~f\left(\boldsymbol{u} \right).
\end{equation}

In this paper, the optimization function \eqref{eq:3} will be called \ctp{Combined} (\ctp{C}), and \eqref{eq:2} will be called \ctp{Energy} (\ctp{E}). To distinguish the optimization strategies and functions, we use the following abbreviation scheme. The first letter denotes the type of optimization function (\ctp{E} or \ctp{C}). The second letter stands for the applied optimization strategy, denoted by \ctp{1} if only one tool can be switched, or by \ctp{A} if all coding tools can be changed. In summary, this corresponds to the following four optimization strategies that will be evaluated to derive optimal CTPs: \ctp{EA}, \ctp{E1}, \ctp{CA}, \ctp{C1}. Strategy \ctp{EA} corresponds to the method that was originally proposed in~\cite{Kraenzler21}.

\section{Setup and Metrics}
\label{sec:Setup}

For the evaluation of the DSE, we use two practical implementations of VVC, which are VVenC~\cite{VVenC} for encoding of bit streams and VVdeC~\cite{VVdeC} for decoding. For VVenC and VVdeC, we both use version 1.30 of the repositories in \cite{VVenC130} and \cite{VVdeC130}, respectively.

To measure the objective visual quality of the encoded bit streams, we both use PSNR~\cite{WorkingPractices} and VMAF. For the calculation of VMAF, we use version 2.3.0 of~\cite{Netflix2022}.

For the measurement of the decoding energy demand, we use a desktop PC with CentOS and an Intel i7-8700 CPU, which has six cores and uses the x86 architecture. Furthermore, the CPU has an integrated power meter that enables power measurements~\cite{DavidGorbatovHanebutteEtAl2010}. To ensure statistical correctness, we conduct multiple measurements and verify those with a confidence interval test that was also used in~\cite{Kraenzler21}.

As test sequence sets, we use two sets, which are the common test condition set of JVET~\cite{JVET-T2010}, for which we used the sequences of classes A1-D, and the test set by the ultra video group (UVG)~\cite{MercatViitanenVanne2020}, which has sequences with 4K and full HD resolution and a frame rate of up to 120 frames per second.

In this paper, we evaluate the same set of coding tools as in~\cite{Kraenzler21}, and additionally evaluate the coding tools motion compensated temporal filtering (MCTF) and transform skip residual coding (TSRC). A detailed description of all 30 coding tools that were evaluated can be found in~\cite{Bross2021a}.

To evaluate the energy and compression efficiency, we use the following Bj{\o}ntegaard-Delta~(BD) metrics based on the descriptions in~\cite{Kraenzler21}: BDDE-PSNR, BDDE-VMAF, BDR-PSNR, and BDR-VMAF. For the evaluation of the compression efficiency, we use  Bj{\o}ntegaard-Delta Bit Rate (BDR) to evaluate the bit rate increase or decrease for the same objective visual quality in terms of PSNR or VMAF. Accordingly, we use Bj{\o}ntegaard-Delta Decoding Energy (BDDE) to evaluate energy efficiency for the same visual quality level. Throughout this work, the \ctp{slower} preset of VVenC will be the reference for the BD-metric calculations.

\begin{center}

\begin{figure*}[htbp]
\centering
\vspace{-0.2cm}
\begin{tikzpicture}
\begin{axis}[
 width=0.9\textwidth,
 height = 0.5\textwidth,
    xlabel={BDR-VMAF in $\%$},
    ylabel={BDDE-VMAF in $\%$},
legend cell align = {left},
    xmin=-2, xmax=42,
    ymin=-45, ymax=5,
        axis lines = left,
    ymajorgrids=true,
    xmajorgrids=true,
    grid style=dashed,
        yshift=0.33cm,
]

\definecolor{EA}{HTML}{00D66D}
\definecolor{CA}{HTML}{EAF500}
\definecolor{EB}{HTML}{F56E00}
\definecolor{CB}{HTML}{D6003B}
\definecolor{ETT}{HTML}{0012EB}
\definecolor{CTT}{HTML}{627380}
\definecolor{Gray}{HTML}{808080}

\addplot[only marks,
   color=black,
    mark=*,    
    mark size=1.5pt,
    line width=0.25pt,
    ]
    coordinates {
(4.224,-1.49) (7.886,-11.44) (1.205,1.244) (0.264,0.9198) (1.214,-3.144) (-0.0625,0.2481) (-0.1056,0.4848) (0.2904,-0.03215) (0.2489,-0.05702) (0.2435,0.8046) (1.774,-6.255) 
(1.774,-6.255) (0.9427,-3.339) (1.77,1.186) (0.6154,0.463) (-0.1104,-4.329) (0.0133,-0.1964) (0.1345,0.1249) (0.7357,0.2182) (0.3506,-2.236) (0.3494,-0.1352) (0.6367,0.9046) 
(0.6367,0.9046) (0.3481,0.06012) (0.5436,0.292) (0.5829,-0.611) (0.1221,-0.2739) (0.6207,-0.1465) (0.3702,0.5926) (0.1482,0.358) (7.362,1.194) (0.1311,-0.04341) 
    };
    \addlegendentry{1.Iteration}

\addplot[only marks,
   color=black,
   fill=EA,
    mark=*,
    mark size=1.5pt,
    line width=0.25pt,
    ]
    coordinates {
(20.59,-39.02) (23.38,-39.83) (23.33,-39.14) (23.11,-39.89) (23.06,-37.82) (23.38,-39.16) (20.76,-37.25) (12.31,-29.69) (14.3,-31.49) (14.31,-30.59) (14.35,-31.36) 
(14.35,-31.36) (16.07,-30.61) (14.9,-30.45) (14.27,-30.04) (25.81,-39.46) (28.8,-40.74) (28.75,-40.11) (28.4,-40.46) (31.49,-39.99) (28.58,-40.05) (28.79,-39.6) 
(28.79,-39.6) (25.3,-40.31) (28.33,-41.33) (28.27,-40.64) (27.63,-41.28) (30.15,-41.05) (28,-40.64) (27.42,-40.5) (11.81,-36.63) (12.7,-35.42) (12.66,-34.84) 
(12.66,-34.84) (12.47,-35.59) (17.61,-37.52) (12.52,-35.02) (15.14,-36.51) (25.14,-40.46) (28.3,-41.18) (28.18,-40.81) (27.63,-41.35) (30.18,-41.12) (27.9,-41.05) 
(27.9,-41.05) (27.73,-40.33) (25.14,-40.53) (28.43,-41.1) (28.18,-40.86) (27.63,-41.38) (30.29,-41.13) (27.9,-40.61) (27.73,-40.45) (24.84,-40.24) (28.12,-40.74) 
(28.12,-40.74) (27.86,-40.31) (27.69,-40.9) (29.91,-40.85) (27.71,-40.35) (27.18,-39.91) (25.16,-39.91) (28.23,-40.37) (28.02,-40.11) (27.61,-40.14) (30.38,-40.46) 
(30.38,-40.46) (27.87,-40.18) (27.76,-39.71) (25.17,-40.47) (27.93,-41.23) (28.21,-40.62) (27.65,-40.02) (30.04,-41.1) (27.96,-40.79) (27.78,-40.07) (23.28,-36.31) 
(23.28,-36.31) (26.23,-37.46) (26.26,-37.08) (26.29,-36.59) (28.23,-36.55) (26.26,-37.04) (25.89,-35.57) (21.09,-36) (23.2,-35.41) (23.07,-35.26) (23.13,-34.72) 
(23.13,-34.72) (24.76,-35.89) (23.28,-35.35) (22.46,-34.86) (27.15,-39.94) (30.58,-40.09) (30.42,-40.11) (29.89,-39.54) (32.52,-40.54) (30.22,-40.28) (29.88,-39.4) 
(29.88,-39.4) (26.41,-40.72) (26.65,-40.88) (26.6,-41.24) (29.15,-40.33) (31.76,-41.02) (29.53,-40.68) (28.98,-40.21) (25.39,-36.83) (28.43,-36.47) (28.29,-36.88) 
(28.29,-36.88) (28.06,-36.09) (30.52,-37.46) (28.11,-36.86) (28.07,-36.48) (24.77,-39.95) (27.83,-40.39) (28.02,-40.26) (27.27,-40.06) (29.62,-41.11) (27.79,-41.09) 
(27.79,-41.09) (28.01,-40.75) (25.11,-40.33) (27.95,-40.75) (28.04,-40.96) (28,-40.25) (30.41,-41.36) (27.63,-41.56) (27.98,-40.17) (26.51,-40.08) (29.56,-41.04) 
(29.56,-41.04) (29.47,-41.09) (26.46,-39.96) (27.76,-40.85) (26.68,-40.91) (29.76,-40.87) (25.36,-37.55) (28.8,-38.74) (28.28,-38.85) (27.94,-38.23) (30.56,-39.28) 
(30.56,-39.28) (28.31,-39.36) (28.09,-38.37) (24.83,-39.15) (27.65,-40.26) (27.91,-40.21) (27.37,-39.82) (29.57,-40.87) (27.62,-40.58) (27.39,-39.97) (27,-39.8) 
(27,-39.8) (30.57,-40.89) (30.53,-40.8) (30.27,-41.45) (28.32,-41.58) (30.53,-41.3) (25.81,-40.52) (25.21,-39.79) (28.45,-40.9) (28.28,-40.89) (28.05,-41.47) 
(28.05,-41.47) (29.83,-41.36) (28.41,-41.26) (27.59,-40.49) (26.2,-39.73) (29.28,-40.62) (29.3,-40.58) (29.65,-41.65) (27.9,-41.16) (30.19,-41.44) (26.22,-40.66) 
(26.22,-40.66) (25.14,-39.92) (28.43,-40.96) (28.18,-40.9) (27.63,-41.38) (30.29,-41.43) (27.9,-41.45) (27.73,-40.54) (27.14,-37.57) (30.43,-38.87) (30.22,-38.66) 
(30.22,-38.66) (29.64,-39.41) (32.75,-39.34) (30.02,-39.24) (29.95,-38.36) (24.45,-40.15) (27.68,-41.1) (27.31,-40.82) (27.1,-41.43) (30.81,-41.36) (27.23,-41.18) 
(27.23,-41.18) (27.9,-40.25) (26.41,-41.01) (26.59,-40.66) (26.54,-40.39) (26.27,-41.19) (31.7,-41.73) (26.68,-41.04) (29.18,-40.83) (25.19,-40.55) (28.42,-40.93) 
(28.42,-40.93) (28.09,-40.81) (27.83,-41.59) (30.63,-41.48) (27.99,-41.29) (28.11,-40.52) (35.11,-39.28) (38.67,-39.43) (38.4,-39.35) (37.93,-40.13) (40.42,-40.07) 
(40.42,-40.07) (38.27,-39.82) (37.74,-38.97) (25.12,-40.78) (28,-41.23) (28.43,-40.84) (27.85,-41.38) (30.22,-41.24) (28,-41.28) (27.59,-40.61) 
    };
    \addlegendentry{\ctp{EA}}

\addplot[only marks,
   color=black,
   fill=EB,
    mark=*,
    mark size=1.5pt,
    line width=0.25pt,
    ]
    coordinates {
(10.76,-13.28) (13.33,-19.66) (13.03,-24.61) (21.65,-27.88) (25.6,-34.04) (19.66,-30.86) (19.5,-33.8) (17.73,-36.07) (17.64,-36.36) (18.72,-37.51) (19.09,-36.24) 
(19.09,-36.24) (19.41,-36.76) (19.76,-36.94) (21,-38.01) (23.1,-38.4) (23.1,-37.6) (23.34,-38.27) (0,-0.1726) (1.774,-6.933) (1.576,-11.73) (2.616,-13.05) 
(2.616,-13.05) (4.211,-19.42) (9.986,-24.33) (10.5,-27.26) (10.67,-27.96) (10.98,-28.97) (11.7,-31.47) (12.07,-29.71) (12.35,-29.4) (12.46,-30.31) (13.5,-31.4) 
(13.5,-31.4) (14.93,-32.08) (14.59,-31.44) (14.68,-31.47) (8.706,-10.21) (10.86,-17.45) (10.96,-22.27) (17.42,-25.03) (20.35,-29.68) (26.61,-33.13) (26.64,-35.94) 
(26.64,-35.94) (24.47,-37.98) (24.4,-38.58) (25.87,-39.97) (26.03,-38.97) (26.25,-39.24) (26.67,-39.92) (27.98,-40.75) (30.22,-41.07) (30.22,-40.59) (30.17,-40.77) 
(30.17,-40.77) (7.492,-11.42) (9.914,-17.66) (9.875,-22.96) (16.63,-26.14) (19.69,-30.7) (25.41,-33.97) (25.51,-36.85) (23.78,-38.77) (23.52,-39.47) (25.03,-40.63) 
(25.03,-40.63) (25.28,-39.89) (25.4,-40.21) (25.73,-40.48) (27.39,-41.46) (29.31,-41.48) (29.31,-41.16) (29.38,-41.43) (14.26,-15.6) (16.81,-22.71) (16.61,-27.99) 
(16.61,-27.99) (9.978,-21.88) (10.1,-27.03) (13.14,-29.12) (13.48,-32.28) (11.04,-34.79) (10.91,-35.46) (11.33,-35.98) (11.51,-34.85) (11.93,-35.75) (11.84,-35.42) 
(11.84,-35.42) (13.22,-36.48) (14.93,-36.71) (14.93,-36.04) (14.86,-36.8) (7.679,-11.62) (10.16,-18.12) (9.818,-23.24) (16.66,-26.26) (19.53,-30.89) (25.32,-34.22) 
(25.32,-34.22) (25.48,-37.18) (23.68,-39.02) (23.65,-39.43) (24.99,-40.82) (25.25,-39.62) (25.66,-40.83) (25.74,-40.56) (26.92,-41.59) (29.24,-41.72) (29.24,-41.2) 
(29.24,-41.2) (29.21,-41.45) (7.886,-11.36) (10.16,-18.22) (9.978,-23.35) (16.61,-26.4) (19.66,-30.93) (25.6,-34.02) (25.48,-36.79) (23.76,-38.89) (23.58,-39.42) 
(23.58,-39.42) (25.1,-40.8) (25.22,-39.52) (25.48,-41.03) (25.78,-40.42) (27.25,-41.58) (29.08,-41.85) (29.08,-41.16) (29.21,-41.41) (7.875,-12.17) (10.35,-18.87) 
(10.35,-18.87) (10.13,-23.68) (17.01,-26.63) (19.89,-31.35) (25.65,-34.7) (25.62,-37.77) (24.07,-39.24) (23.92,-40.1) (25.22,-41.41) (25.1,-39.07) (25.48,-40.29) 
(25.48,-40.29) (25.6,-40.04) (26.9,-41.03) (29.06,-41.41) (29.06,-40.76) (28.91,-41.22) (7.815,-12.14) (10.15,-18.88) (10,-23.83) (16.62,-26.68) (19.68,-31.23) 
(19.68,-31.23) (25.46,-34.98) (25.38,-37.85) (23.58,-39.48) (23.76,-38.95) (24.99,-40.04) (25.4,-38.9) (25.47,-40.34) (25.71,-39.83) (27.13,-40.87) (29.19,-41.07) 
(29.19,-41.07) (29.19,-40.34) (29.1,-40.8) (8.012,-11.06) (10.17,-17.97) (9.978,-23.02) (17.01,-26.08) (19.89,-30.43) (25.67,-33.89) (25.63,-36.89) (23.82,-38.47) 
(23.82,-38.47) (23.86,-39.11) (25.18,-40.44) (25.22,-39.39) (25.81,-40.86) (25.96,-40.25) (27.16,-41.6) (29.26,-41.63) (29.26,-41.08) (29.21,-41.41) (10.16,-17.95) 
(10.16,-17.95) (7.886,-11.65) (7.473,-16.45) (14.08,-19.05) (16.72,-24.06) (22.4,-29.72) (22.5,-32.46) (22.2,-33.95) (22.14,-34.61) (23.48,-37.2) (23.66,-35.7) 
(23.66,-35.7) (24.18,-37.14) (24.33,-36.87) (25.59,-37.97) (27.53,-38.05) (27.53,-37.35) (27.31,-37.85) (7.818,-15.67) (10.21,-22.94) (10.1,-27.98) (19.66,-30.47) 
(19.66,-30.47) (16.61,-27.36) (21.65,-29.07) (21.52,-31.67) (19.62,-34.14) (19.52,-34.71) (20.4,-35.89) (20.65,-34.72) (21.01,-35.57) (21.17,-35.4) (22.58,-36.08) 
(22.58,-36.08) (24.62,-36.66) (24.62,-35.83) (24.39,-36.78) (9.626,-10.61) (12.07,-17.44) (11.85,-22.63) (18.61,-25.19) (21.6,-30) (27.33,-32.98) (27.36,-36.17) 
(27.36,-36.17) (25.73,-38.11) (25.5,-38.56) (26.93,-40.06) (27.29,-39.12) (27.57,-40.15) (27.91,-39.69) (29.29,-40.62) (31.55,-40.98) (31.55,-40.51) (31.5,-41.24) 
(31.5,-41.24) (8.658,-10.81) (11,-18.21) (10.57,-23.11) (17.7,-25.92) (20.71,-30.56) (26.66,-33.8) (26.59,-36.85) (24.86,-38.59) (24.82,-39.44) (26.4,-40.54) 
(26.4,-40.54) (26.73,-39.98) (26.97,-40.75) (27.26,-40.3) (28.52,-41.63) (30.64,-41.66) (30.64,-41.12) (30.69,-41.34) (7.473,-16.06) (9.978,-24.09) (10.16,-18.38) 
(10.16,-18.38) (16.81,-20.9) (19.84,-26.26) (25.74,-29.49) (25.58,-32.54) (23.91,-34.36) (23.88,-35.12) (25.1,-36.63) (25.4,-35.94) (25.73,-36.8) (25.91,-36.52) 
(25.91,-36.52) (27.4,-37.47) (29.48,-37.83) (29.48,-37.35) (29.27,-37.56) (7.692,-11.91) (10.35,-19.1) (10.14,-23.57) (16.92,-26.56) (19.76,-31.24) (25.84,-34.28) 
(25.84,-34.28) (25.48,-37.45) (23.91,-38.99) (23.97,-39.54) (25.19,-40.87) (25.4,-40.37) (25.78,-41.06) (25.48,-40.16) (26.71,-41.22) (28.85,-41.31) (28.85,-40.85) 
(28.85,-40.85) (28.92,-41.26) (7.942,-11.43) (10.27,-19.25) (10.13,-23.07) (17.09,-26.98) (19.92,-30.97) (25.67,-33.97) (25.58,-37.23) (23.93,-38.68) (23.78,-39.33) 
(23.78,-39.33) (25.35,-40.66) (25.49,-39.94) (25.92,-40.75) (26.02,-40.36) (27.45,-41.37) (29.73,-41.8) (29.73,-41.16) (29.59,-40.53) (8.706,-11.46) (10.93,-19.18) 
(10.93,-19.18) (10.76,-23.3) (17.6,-27.7) (20.51,-31.53) (26.43,-34.3) (26.63,-37.14) (24.77,-38.98) (24.59,-39.6) (26.06,-40.87) (26.32,-40.8) (26.71,-41.04) 
(26.71,-41.04) (27.25,-40.91) (25.78,-41.19) (27.09,-40.98) (27.09,-40.69) (27.12,-39.98) (7.856,-13.56) (10.18,-21.17) (9.904,-25.63) (16.57,-29.78) (19.5,-33.2) 
(19.5,-33.2) (25.48,-36.17) (25.6,-35.06) (23.82,-36.71) (23.61,-37.49) (25.16,-38.72) (25.39,-39.05) (25.79,-38.73) (26.06,-38.45) (27.44,-39.64) (29.26,-39.73) 
(29.26,-39.73) (29.26,-39.24) (29.47,-38.47) (8.17,-10.82) (10.59,-19.45) (10.56,-23.75) (17.31,-28.12) (20.15,-31.7) (26.02,-34.62) (25.93,-37.51) (24.08,-39.3) 
(24.08,-39.3) (24.08,-40.04) (25.48,-41.33) (25.48,-41.43) (25.22,-40.31) (25.4,-40.19) (26.61,-40.97) (28.59,-41.07) (28.59,-40.71) (28.91,-39.76) (9.425,-10.76) 
(9.425,-10.76) (11.79,-18.19) (11.74,-22.58) (17.2,-26.85) (20.62,-30.68) (27.34,-34.14) (27.5,-36.84) (25.69,-38.84) (25.4,-39.35) (26.87,-40.62) (27.05,-41.08) 
(27.05,-41.08) (27.35,-40.67) (27.6,-40.15) (29.09,-41.46) (31.13,-41.74) (31.13,-41.22) (31.22,-40.46) (8.039,-11.49) (10.32,-18.69) (10.16,-23.16) (16.71,-27.57) 
(16.71,-27.57) (19.71,-31.17) (25.83,-34.19) (25.57,-37.13) (23.77,-38.85) (23.8,-39.57) (25.13,-40.67) (25.62,-40.93) (25.83,-40.57) (26.14,-40.04) (27.57,-41.4) 
(27.57,-41.4) (29.55,-41.44) (29.55,-41.72) (29.49,-40.37) (8.499,-11.62) (11.45,-18.86) (11.24,-23.28) (18.18,-27.3) (21.05,-30.96) (26.98,-34.08) (26.79,-36.79) 
(26.79,-36.79) (25.06,-38.69) (25.04,-39.44) (26.39,-40.65) (26.49,-41) (27.06,-40.56) (27.09,-40.02) (29.08,-41.76) (27.25,-41.55) (27.25,-41.32) (27.03,-40.28) 
(27.03,-40.28) (8.524,-12.48) (11.08,-19.65) (10.77,-24.17) (17.79,-28.72) (20.8,-32.84) (25.6,-34) (25.48,-37.08) (23.76,-38.89) (23.58,-40.16) (25.1,-40.71) 
(25.1,-40.71) (25.22,-41.23) (25.48,-40.88) (25.78,-40.36) (27.25,-41.58) (29.08,-41.69) (29.08,-41.56) (29.21,-40.48) (6.817,-12.91) (7.99,-20.86) (7.749,-25.32) 
(7.749,-25.32) (14.86,-29.71) (17.88,-33.25) (23.82,-36.04) (23.76,-38.81) (25.48,-36.86) (25.38,-38.7) (26.85,-38.95) (27.08,-39.24) (27.55,-38.97) (27.85,-38.41) 
(27.85,-38.41) (29.15,-39.53) (31.52,-39.72) (31.52,-39.57) (31.71,-38.3) (8.212,-11.27) (10.52,-18.64) (10.33,-23.08) (17.25,-27.46) (20.27,-31.03) (26.2,-34.38) 
(26.2,-34.38) (26.33,-37.02) (24.27,-38.88) (24.11,-40.48) (25.64,-40.51) (26.02,-41.22) (26.04,-40.73) (26.51,-40.52) (27.83,-41.74) (29.63,-41.71) (29.63,-41.58) 
(29.63,-41.58) (29.67,-40.51) (8.283,-10.99) (10.5,-17.71) (10.43,-21.92) (17.53,-26.94) (21,-30.72) (26.67,-34.77) (26.82,-37.38) (24.99,-39.23) (25.1,-40.7) 
(25.1,-40.7) (23.58,-40.28) (23.92,-39.93) (23.95,-40.51) (24.45,-40.26) (25.84,-41.26) (27.88,-41.5) (27.88,-41.18) (27.81,-40.05) (7.686,-12.3) (10.03,-18.59) 
(10.03,-18.59) (9.846,-22.77) (16.71,-27.56) (19.72,-30.98) (25.56,-34.26) (25.47,-37.07) (23.66,-38.98) (23.59,-40.24) (24.86,-40.66) (25.13,-40.2) (25.52,-40.93) 
(25.52,-40.93) (25.68,-40.75) (27.14,-41.52) (29.07,-41.57) (29.07,-41.68) (29.08,-40.39) (16.26,-10.03) (18.6,-17.34) (18.67,-21.68) (26.99,-25.95) (30.33,-29.52) 
(30.33,-29.52) (36.35,-32.85) (36.3,-35.39) (33.45,-37.38) (33.38,-38.94) (34.84,-39.16) (34.9,-38.71) (35.42,-39.34) (35.91,-39.3) (37,-40.28) (39.6,-40.29) 
(39.6,-40.29) (39.6,-40.24) (39.6,-39.14) (7.842,-11.3) (9.883,-18.8) (9.716,-23.24) (16.45,-27.79) (19.44,-31.25) (25.44,-34.41) (25.28,-37.01) (23.72,-39.03) 
(23.72,-39.03) (23.65,-40.31) (24.91,-40.66) (25.28,-40.19) (25.7,-40.96) (25.86,-40.84) (27.03,-41.71) (29.21,-41.65) (29.21,-41.77) (29.08,-40.56) 
    };
    \addlegendentry{\ctp{E1}}

\addplot[only marks,
 color=black,
    fill=CA,
    mark=*,
    mark size=1.5pt,
    line width=0.25pt,
    ]
    coordinates {
(23.91,-39.16) (8.634,-22.77) (24.56,-40.58) (8.758,-22.33) (23.83,-38.97) (8.404,-23.02) (24.35,-39.41) (7.951,-23.1) (24.04,-39.8) (9.533,-22.26) (24.44,-40.19) 
(24.44,-40.19) (7.889,-22.67) (5.469,-26) (9.105,-35.47) (6.607,-27.3) (9.514,-35.35) (5.558,-25.92) (8.809,-34.31) (5.779,-26.49) (9.678,-35.17) (6.089,-26.41) 
(6.089,-26.41) (8.635,-34) (5.931,-26.93) (9.68,-35.12) (19.98,-36.96) (5.918,-20.51) (20.31,-38.53) (6.049,-20.83) (19.67,-36.73) (4.954,-19.6) (20.6,-37.2) 
(20.6,-37.2) (5.086,-20.98) (20.08,-37.94) (6.182,-18.91) (20.34,-38.16) (4.915,-20.48) (19.34,-37.86) (4.561,-21.78) (19.98,-39.33) (4.76,-21.25) (19.17,-37.66) 
(19.17,-37.66) (3.966,-20.58) (20.09,-37.76) (3.953,-21.9) (19.78,-39.04) (4.869,-20.06) (19.87,-38.88) (4.045,-21.29) (9.243,-34.23) (6.674,-27.94) (9.549,-35.24) 
(9.549,-35.24) (6.716,-27.87) (8.75,-33.65) (5.595,-25.61) (9.678,-34.16) (5.779,-27.48) (9.616,-35.33) (6.907,-25.88) (9.68,-35.19) (5.931,-27.14) (19.41,-37.98) 
(19.41,-37.98) (4.473,-21.52) (19.81,-39.21) (4.66,-21.71) (19.13,-37.64) (3.497,-20.71) (19.86,-37.77) (3.861,-21.83) (19.64,-39.03) (4.775,-20.06) (19.76,-38.73) 
(19.76,-38.73) (3.958,-21.44) (19.37,-38.12) (4.768,-21.37) (19.81,-39.28) (4.498,-21.97) (19.13,-37.71) (3.999,-20.6) (19.86,-37.77) (4.031,-21.35) (19.64,-38.83) 
(19.64,-38.83) (4.522,-20.25) (19.67,-38.2) (4.152,-21.1) (19.51,-38.46) (4.711,-21.23) (19.81,-38.8) (4.58,-21.22) (18.83,-37.08) (3.64,-20.16) (19.59,-36.82) 
(19.59,-36.82) (3.685,-21.31) (19.34,-38.16) (4.669,-19.6) (19.45,-37.97) (3.69,-21.13) (19.3,-38.62) (4.223,-20.9) (20.02,-38.61) (4.329,-21.19) (19.19,-38.06) 
(19.19,-38.06) (3.358,-20.15) (19.8,-37.32) (4.042,-22.55) (19.62,-38.06) (4.4,-19.62) (19.88,-38.88) (4.315,-21.96) (19.2,-37.7) (4.628,-21.34) (19.73,-39.28) 
(19.73,-39.28) (4.725,-21.14) (19.26,-37.39) (3.662,-20.54) (19.78,-37.14) (3.74,-21.97) (19.66,-38.48) (4.865,-19.57) (19.76,-38.24) (3.962,-21.11) (17.99,-31.35) 
(17.99,-31.35) (2.65,-13.24) (18.06,-32.5) (3.079,-13.7) (17.63,-30.66) (2.093,-12.95) (18.37,-30.21) (2.215,-14.02) (18.22,-32.13) (3.077,-11.83) (18.22,-31.65) 
(18.22,-31.65) (2.228,-13.73) (16.14,-33.98) (3.694,-16.69) (16.71,-34.9) (3.868,-17.09) (16,-33.71) (3.036,-16.85) (16.63,-32.87) (2.966,-17.74) (16.32,-34.57) 
(16.32,-34.57) (3.916,-15.85) (16.61,-34.06) (2.973,-17.33) (21.55,-37.15) (6.665,-19.93) (22.28,-38.35) (6.56,-20.8) (21.58,-37.03) (5.792,-19.51) (22.04,-36.47) 
(22.04,-36.47) (5.53,-21.14) (21.89,-37.94) (7.094,-18.78) (22.05,-37.51) (5.787,-20.67) (20.46,-37.69) (5.12,-20.69) (21.01,-39.01) (5.457,-21.36) (20.3,-37.88) 
(20.3,-37.88) (4.528,-20.51) (20.96,-36.96) (4.469,-21.55) (20.67,-38.48) (5.504,-19.18) (20.73,-37.96) (4.521,-21.25) (19.33,-33.87) (4.553,-16.22) (20.07,-34.89) 
(20.07,-34.89) (4.804,-17.14) (19.37,-33.55) (3.721,-16.17) (20.01,-32.43) (4.031,-17.25) (19.56,-34.3) (4.969,-15.3) (20.29,-34.08) (4.06,-17.21) (19.28,-37.7) 
(19.28,-37.7) (4.525,-20.54) (19.61,-38.72) (4.377,-21.25) (19.35,-38.33) (3.615,-20.47) (19.96,-37.37) (3.853,-22.54) (19.54,-38.31) (4.73,-19.46) (19.66,-37.94) 
(19.66,-37.94) (3.81,-21.3) (19.75,-38) (4.472,-20.85) (19.99,-39.24) (4.771,-21.2) (19.5,-37.76) (3.973,-20.73) (19.43,-37.32) (3.709,-21.84) (19.74,-38.63) 
(19.74,-38.63) (5.066,-19.45) (20.08,-37.97) (4.147,-21.23) (20.52,-38.42) (5.568,-21.05) (20.89,-39.53) (5.589,-21.73) (20.12,-38.15) (4.605,-20.96) (20.85,-37.31) 
(20.85,-37.31) (4.689,-21.76) (20.95,-38.94) (5.875,-19.86) (20.94,-38.13) (4.879,-21.69) (19.57,-35.86) (4.253,-18.37) (20.25,-36.83) (4.45,-19.22) (19.42,-35.59) 
(19.42,-35.59) (3.378,-18.28) (20.24,-34.58) (3.558,-19.06) (19.85,-36.24) (4.528,-17.21) (20.14,-35.85) (3.478,-19.17) (19.5,-38.76) (4.052,-20.4) (19.44,-38.57) 
(19.44,-38.57) (4.366,-21.18) (19.49,-38.42) (3.35,-20.14) (20.15,-37.77) (4.11,-21.21) (20.03,-39.17) (4.408,-19.08) (19.61,-37.57) (3.605,-21.03) (20.18,-37.93) 
(20.18,-37.93) (5.654,-20.72) (20.53,-38.76) (5.788,-21.32) (20.01,-37.47) (4.481,-20.5) (20.78,-37.02) (4.665,-20.09) (20.48,-38.48) (5.932,-19.24) (20.6,-37.66) 
(20.6,-37.66) (4.658,-21.39) (19.67,-37.98) (4.858,-20.84) (20.05,-39.06) (4.988,-21.27) (19.35,-37.94) (3.816,-20.84) (19.71,-37.34) (3.722,-20.22) (19.33,-38.99) 
(19.33,-38.99) (4.844,-19.52) (20.05,-38.13) (4.272,-21.33) (20.79,-38.03) (5.38,-20.69) (21.36,-38.86) (5.635,-21.14) (20.52,-37.91) (4.403,-20.32) (21.04,-37.13) 
(21.04,-37.13) (4.674,-20.01) (20.9,-39.77) (5.281,-20.16) (21.17,-37.88) (4.733,-21.41) (18.14,-36.54) (3.987,-19.44) (18.46,-37.49) (4.125,-19.96) (17.61,-36.3) 
(17.61,-36.3) (4.19,-21.74) (18.5,-35.6) (3.269,-18.85) (18.09,-38.29) (5.296,-21.31) (18.3,-36.48) (3.189,-20.26) (20.94,-36.07) (4.552,-19.61) (21.65,-36.8) 
(21.65,-36.8) (4.712,-20.14) (20.81,-35.84) (3.795,-19.65) (21.48,-35.12) (3.709,-19.55) (21.37,-37.81) (4.948,-18.71) (21.29,-35.96) (3.862,-20.41) (19.93,-38.14) 
(19.93,-38.14) (5.15,-20.91) (20.6,-39.08) (5.382,-20.19) (19.79,-37.87) (4.423,-20.93) (20.44,-37.42) (4.576,-20.61) (20.37,-39.67) (4.212,-19.78) (20.46,-38.11) 
(20.46,-38.11) (4.461,-21.36) (20.57,-38.47) (4.899,-19.67) (20.92,-39.08) (5.163,-18.99) (20.28,-37.93) (4.19,-18.96) (21.08,-37.96) (4.342,-20.42) (20.81,-39.77) 
(20.81,-39.77) (5.352,-18.31) (21.08,-37.91) (4.291,-20.22) (19.36,-38.29) (4.293,-21.02) (19.93,-39.1) (4.396,-20.56) (19.11,-37.91) (3.921,-20.34) (19.72,-37.77) 
(19.72,-37.77) (4.025,-21.23) (19.42,-39.69) (4.537,-20.1) (19.69,-38.06) (4.075,-21.06) (29.33,-36.85) (12.34,-19.27) (29.49,-37.79) (12.24,-18.76) (28.79,-36.45) 
(28.79,-36.45) (11.45,-19.44) (29.62,-36.43) (11.11,-20.14) (29.49,-38.5) (12.82,-18.61) (29.57,-36.69) (11.58,-20.11) (19.41,-38.12) (4.463,-20.7) (19.82,-39.06) 
(19.82,-39.06) (4.706,-20.15) (19.2,-38.02) (3.594,-20.71) (19.82,-37.94) (3.885,-21.85) (19.59,-39.58) (4.865,-19.87) (19.78,-38.18) (3.876,-21.62) 
    };
    \addlegendentry{\ctp{CA}}

\addplot[only marks,
   color=black,
   fill=CB,
    mark=*,
    mark size=1.5pt,
    line width=0.25pt,
    ]
    coordinates {
(6.082,-7.919) (5.999,-14.72) (7.755,-17.87) (9.986,-25.67) (9.986,-25.03) (10.5,-26.36) (10.67,-28.19) (10.98,-28.01) (11.09,-29.51) (10.94,-32.01) (11.02,-30.33) 
(11.02,-30.33) (10.81,-29.74) (10.16,-18.04) (9.978,-25.05) (16.61,-28.31) (19.66,-31.77) (20.8,-32.82) (20.71,-35.17) (19.28,-37.63) (18.92,-37.61) (19.33,-39.01) 
(19.33,-39.01) (19.33,-40.29) (19.13,-39.24) (19.24,-39.13) (2.916,-6.853) (2.726,-12.66) (3.886,-15.21) (5.421,-19.48) (5.997,-21.23) (6.5,-22.89) (6.679,-24.59) 
(6.679,-24.59) (6.879,-24.16) (7.224,-26.13) (6.947,-28.37) (6.875,-26.71) (6.801,-26.42) (2.083,-6.496) (1.749,-13.17) (2.752,-16.29) (4.414,-20.5) (5.025,-21.91) 
(5.025,-21.91) (5.632,-23.85) (5.595,-25.65) (5.884,-25.37) (6.205,-26.88) (5.943,-29.3) (5.939,-27.36) (5.878,-27.3) (2.939,-10.79) (2.616,-17.73) (1.576,-12.4) 
(1.576,-12.4) (2.381,-15.06) (2.917,-16.02) (3.362,-18.5) (3.468,-20.22) (3.66,-19.96) (4.031,-21.44) (3.948,-23.72) (3.797,-22.02) (3.635,-21.81) (1.824,-7.018) 
(1.824,-7.018) (1.555,-13.61) (2.778,-16.24) (4.226,-20.51) (4.806,-21.65) (5.225,-23.95) (5.515,-25.37) (5.553,-25.34) (6.049,-26.86) (5.748,-29.42) (5.794,-27.31) 
(5.794,-27.31) (5.686,-27.07) (1.609,-7.301) (1.358,-13.41) (2.56,-16.08) (4.073,-20.36) (4.696,-21.77) (5.014,-24.17) (5.318,-25.44) (5.606,-25.32) (5.794,-27.11) 
(5.794,-27.11) (6.026,-29.27) (6.049,-27.29) (5.97,-27.06) (2.057,-7.253) (1.869,-14.06) (3.041,-16.71) (4.588,-21.01) (5.277,-22.02) (5.648,-24.35) (5.806,-25.97) 
(5.806,-25.97) (6.026,-26.96) (5.762,-26.41) (5.606,-29.12) (5.631,-26.78) (5.507,-26.72) (2.121,-7.153) (1.884,-13.98) (3.063,-16.52) (4.645,-21.13) (5.105,-21.77) 
(5.105,-21.77) (5.597,-24.39) (5.762,-26.32) (5.425,-25.16) (5.806,-26.52) (5.631,-28.84) (5.486,-26.93) (5.259,-26.52) (2.023,-6.774) (1.784,-13.31) (2.992,-16.02) 
(2.992,-16.02) (4.451,-20.14) (5.212,-21.18) (5.512,-24.32) (5.614,-25.45) (5.937,-24.91) (6.044,-26.65) (5.804,-29.26) (5.71,-27.7) (5.636,-26.76) (0,-0.4167) 
(0,-0.4167) (-0.1104,-6.62) (1.015,-8.715) (2.323,-13.05) (2.981,-13.78) (3.436,-16.9) (3.676,-17.56) (3.969,-17.12) (4.109,-18.89) (3.854,-21.55) (3.842,-20.12) 
(3.842,-20.12) (3.832,-19.41) (2.715,-11.06) (2.381,-17.46) (4.211,-21.37) (2.616,-15.44) (3.235,-15.06) (3.775,-18.86) (3.942,-19.72) (4.247,-19.39) (4.318,-20.33) 
(4.318,-20.33) (4.262,-23.9) (4.197,-21.99) (4.003,-21.42) (3.482,-5.965) (3.424,-12.71) (4.589,-15.3) (6.119,-19.77) (6.636,-20.71) (7.146,-23.79) (7.407,-24.64) 
(7.407,-24.64) (7.544,-24.34) (7.957,-25.16) (7.664,-28.83) (7.514,-26.95) (7.611,-26.45) (2.404,-6.59) (2.17,-13.34) (3.266,-15.55) (4.881,-20.37) (5.606,-21.2) 
(5.606,-21.2) (5.918,-24.2) (6.094,-25.03) (6.534,-25.15) (6.755,-25.84) (6.506,-29.32) (6.41,-27.64) (6.377,-26.66) (1.576,-11.18) (1.774,-9.191) (2.939,-11.4) 
(2.939,-11.4) (4.549,-17.1) (5.129,-16.61) (5.702,-19.98) (5.71,-20.92) (5.976,-20.58) (6.052,-21.53) (5.968,-24.83) (5.962,-23.41) (5.89,-22.58) (1.934,-7.153) 
(1.934,-7.153) (1.693,-13.91) (2.938,-16.23) (4.46,-22.85) (5.037,-21.88) (5.448,-24.68) (5.469,-25.95) (5.917,-25.4) (6.234,-26.54) (5.898,-29.35) (5.898,-28.09) 
(5.898,-28.09) (5.69,-27.27) (1.988,-6.967) (1.818,-13.81) (2.966,-15.81) (4.412,-22.49) (5.157,-21.42) (5.455,-24.59) (5.712,-25.4) (5.893,-25.13) (6.063,-25.86) 
(6.063,-25.86) (5.974,-29.02) (5.98,-27.72) (5.923,-27.08) (2.674,-7.186) (2.49,-13.65) (3.677,-15.92) (5.1,-22.67) (5.768,-21.66) (6.294,-24.84) (6.345,-25.53) 
(6.345,-25.53) (6.654,-25.24) (6.907,-26.5) (6.637,-29.55) (6.667,-28.26) (6.476,-26.96) (2.415,-9.29) (2.059,-16.18) (3.122,-18.16) (4.736,-24.38) (5.257,-23.45) 
(5.257,-23.45) (4.822,-22.9) (5.04,-23.76) (5.418,-23.36) (5.673,-24.23) (5.357,-27.56) (5.245,-26.12) (5.335,-25.1) (2.033,-7.341) (2.179,-14.2) (3.265,-16.29) 
(3.265,-16.29) (4.896,-23.2) (5.426,-22.24) (5.825,-25.04) (5.958,-26.18) (6.377,-25.77) (6.363,-26.47) (6.198,-29.7) (6.16,-28.32) (6.24,-27.44) (2.33,-6.582) 
(2.33,-6.582) (2.135,-13.35) (3.299,-15.37) (5.035,-22.1) (5.536,-21.44) (6.055,-24.37) (6.239,-25.17) (6.465,-25.2) (6.797,-25.71) (6.503,-28.66) (6.434,-27.7) 
(6.434,-27.7) (6.449,-26.68) (2.012,-6.856) (1.737,-13.39) (2.906,-15.29) (4.457,-22.5) (5.261,-21.59) (5.6,-24.6) (5.663,-25.45) (5.885,-26.44) (6.048,-25.8) 
(6.048,-25.8) (5.849,-29.06) (5.778,-27.83) (5.744,-27.07) (2.416,-8.005) (2.322,-13.59) (3.567,-15.37) (5.02,-22.43) (5.782,-21.84) (6.157,-24.63) (6.178,-25.4) 
(6.178,-25.4) (6.488,-26.52) (6.697,-26.18) (6.628,-29.07) (6.486,-27.81) (6.484,-26.96) (2.358,-8.007) (2.067,-14.38) (3.235,-15.71) (4.822,-24.7) (4.211,-19.87) 
(4.211,-19.87) (4.736,-22.89) (4.839,-23.63) (5.129,-24.55) (5.479,-24.11) (5.203,-28.13) (5.138,-26.26) (5.108,-24.98) (1.67,-7.751) (1.67,-14.12) (2.795,-15.48) 
(2.795,-15.48) (4.552,-23.09) (5.04,-22.67) (5.425,-25.63) (5.257,-24.57) (5.597,-25.71) (6.04,-25.01) (5.696,-28.45) (5.604,-27.19) (5.536,-26.12) (2.518,-6.83) 
(2.518,-6.83) (2.074,-13.13) (3.3,-14.44) (4.97,-22.66) (5.602,-22.01) (6.149,-24.52) (5.975,-25.46) (6.455,-26.51) (6.783,-26.09) (6.576,-29.07) (6.421,-28) 
(6.421,-28) (6.36,-26.99) (2.14,-5.666) (2.012,-12.32) (3.267,-13.64) (4.942,-22.1) (5.503,-22.32) (5.844,-24.83) (6.276,-25.76) (6.487,-26.57) (6.717,-26.77) 
(6.717,-26.77) (6.513,-29.29) (6.478,-28.14) (6.111,-27.22) (2.12,-6.727) (1.708,-13.28) (2.961,-14.29) (4.523,-22.43) (5.18,-22.06) (5.605,-24.46) (5.698,-25.47) 
(5.698,-25.47) (6.061,-26.24) (6.398,-26.38) (6.162,-29) (5.984,-27.9) (5.823,-26.83) (9.161,-5.748) (9.273,-12.29) (10.39,-13.42) (12.01,-21.21) (12.6,-20.73) 
(12.6,-20.73) (13.27,-23.26) (13.23,-24.26) (13.52,-25.06) (13.61,-25.21) (13.42,-27.77) (13.45,-26.67) (13.47,-25.55) (1.776,-6.945) (1.519,-13.71) (2.701,-14.98) 
(2.701,-14.98) (4.294,-22.59) (4.849,-22.25) (5.35,-24.75) (5.528,-25.48) (5.68,-26.53) (5.97,-26.65) (5.686,-29.11) (5.686,-28.04) (5.748,-26.85) 
    };
    \addlegendentry{\ctp{C1}}

 \addplot[ 
   color=green,
    mark=none,
    mark size=0pt,
    line width=1pt,
    ] coordinates {     (40.42,-41.85) (29.08,-41.85) (27.83,-41.74) (27.03,-41.71) (26.92,-41.59) (25.48,-41.43) (25.22,-41.41) (25.19,-40.87) (24.99,-40.82) (24.91,-40.66) (24.86,-40.66) 
(24.86,-40.66) (24.56,-40.58) (23.95,-40.51) (23.65,-40.31) (19.33,-40.29) (19.13,-39.24) (18.09,-38.29) (17.61,-37.52) (14.86,-36.8) (11.81,-36.63) (11.33,-35.98) 
(11.33,-35.98) (9.105,-35.47) (8.809,-34.31) (8.635,-34) (6.198,-29.7) (5.748,-29.42) (5.606,-29.12) (5.203,-28.13) (5.138,-26.26) (5.108,-24.98) (4.822,-24.7) 
(4.822,-24.7) (4.736,-24.38) (4.262,-23.9) (3.948,-23.72) (3.853,-22.54) (3.797,-22.02) (3.74,-21.97) (3.709,-21.84) (3.635,-21.81) (3.605,-21.03) (3.497,-20.71) 
(3.497,-20.71) (3.189,-20.26) (3.122,-18.16) (2.966,-17.74) (2.616,-17.73) (2.381,-17.46) (2.059,-16.18) (1.67,-14.12) (1.519,-13.71) (1.358,-13.41) (1.015,-8.715) 
(1.015,-8.715) (-0.1104,-6.62) (-0.1104,1.244) 
};

\end{axis}
\end{tikzpicture}
\caption{ Evaluation of each optimization strategy, which are represented by different colors. Each marker corresponds to an evaluated CTP that is compared to the default encoded CTP of the VVenC \ctp{slower} preset. The horizontal axis shows BDR-VMAF and the vertical axis BDDE-VMAF. Additionally, a Pareto front (green) is shown, which corresponds to the optimal compression efficiency for a given energy efficiency improvement. Furthermore, the CTPs of the first iteration are shown in black since it is the same for each optimization strategy.}
\label{fig:Derivation}
\vspace{-0.45cm}
\end{figure*}
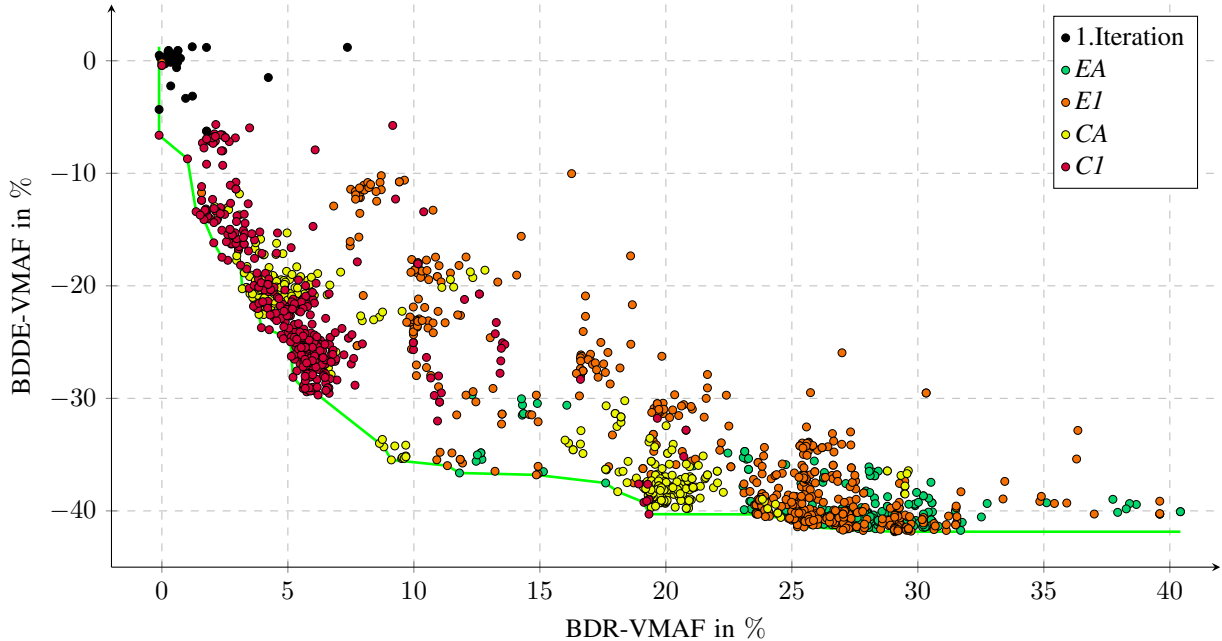
\end{center}

\section{Evaluation}
\label{sec:Evaluation}

\subsection{Design Space Exploration and Derivation of CTPs}

As explained in Section~\ref{sec:Modifications}, we use four optimization strategies to explore our design space. The goals are to find CTPs that have minimum energy demand (\ctp{EE} - Energy Efficient), CTPs with a maximum joint efficiency improvement of energy and bit rate (\ctp{EBE} - Energy and Bit Rate Efficient), or CTPs with a low-bit rate increase and high energy savings (\ctp{LBE} Low Bit Rate Energy Efficient).

For the DSE algorithm, each CTP was encoded with 128 frames for all class B sequences. In Figure~\ref{fig:Derivation}, the results are shown for each optimization strategy with a separate color. Each marker corresponds to a CTP that was compared to the \ctp{slower} preset. Furthermore, a Pareto front is shown (green line) that shows the optimal energy efficiency for a given rate-distortion efficiency. 

For the~\ctp{EA} strategy (green markers), it can be determined that there are no CTPs that have a BDR-VMAF value of less than $10\%$. The lowest BDDE-VMAF value for \ctp{EA} is $\smi41.73\%$. However, the highest energy efficiency among all CTPs was measured for the~\ctp{E1} strategy (orange markers) with a BDDE-VMAF value of $\smi41.85\%$. 
Furthermore, in Figure~\ref{fig:Derivation}, it can be observed that the \ctp{E1} strategy has a broad distribution of CTPs from a BDR-VMAF value of $5\%$ up to $40\%$. However, the rate-distortion efficiency is significantly lower compared to the \ctp{Combined} strategies for low bit rate increases. 

For the \ctp{Combined} optimization strategies, we determine that the Pareto front in Figure~\ref{fig:Derivation} is shifted to the left, which corresponds to less additional bit rate for the same energy savings. For the \ctp{CA} strategy (yellow markers), two main cluster points are surrounded by a large number of CTPs, which are around $5\%$ and $20\%$ BDR-VMAF. Furthermore, there are a few CTPs at $9\%$ BDR-VMAF with a BDDE-VMAF value of $\smi35\%$. 
For the \ctp{C1} strategy (red markers), we determine that most CTPs are between $0\%$ and $7\%$ BDR-VMAF, resulting in a shift of the Pareto front to the left for low bit rate increases. Therefore, energy savings of up to $\smi30\%$ are possible for a bit rate increase of $5\%$. 

\subsection{Validation of the Proposed CTPs}
Based on the results of all optimization strategies, we select for each strategy one CTP with minimum energy demand, which will be called \ctp{EE}, and one CTP with a minimal sum of BDDE and BDR, which will be called \ctp{EBE}. Furthermore, we select four \ctp{LBE} CTPs that have a bit rate increase of less than $5\%$. The usage of the corresponding coding tools for each CTP is given in Table~\ref{tab:CTPUsage}. Furthermore, the evaluation in terms of the BDR and BDDE metrics are shown in Table~\ref{tab:Validation}. In subtable (a), the results for the JVET sequences set are shown and in subtable (b), the corresponding results for the UVG set are given. For comparison, we provide the values for the presets \ctp{slow} and \ctp{medium} of VVenC. Additionally, we show the results for the proposed CTPs in Figure~\ref{fig:JVETValidation} for the JVET set. For reference, the results of the \ctp{EE} and \ctp{EBE} CTP of \cite{Kraenzler21} are also shown. We observe that the \ctp{E1 EE} CTP has the lowest BDDE-VMAF value with $\smi45.31\%$. Furthermore, we show in the figure that each \ctp{EE} CTP has both a higher energy and rate-distortion efficiency than the corresponding profile of~\cite{Kraenzler21}. For the \ctp{EBE}, we have the same observation for the proposed CTPs compared to the \ctp{EBE} CTP of~\cite{Kraenzler21}.

\begin{center}

\begin{figure}[t]
\centering
\vspace{-0.2cm}
\begin{tikzpicture}
\begin{axis}[
 width=0.48\textwidth,
 height = 0.4\textwidth,
    xlabel={BDR-VMAF in $\%$},
    ylabel={BDDE-VMAF in $\%$},
legend cell align = {left},
    xmin=-2, xmax=30,
    ymin=-50, ymax=2,
        axis lines = left,
    ymajorgrids=true,
    xmajorgrids=true,
    grid style=dashed,
  legend columns=2,
        yshift=0.33cm,
]

\definecolor{EA}{HTML}{00D66D}
\definecolor{CA}{HTML}{EAF500}
\definecolor{EB}{HTML}{F56E00}
\definecolor{CB}{HTML}{D6003B}
\definecolor{ETT}{HTML}{0012EB}
\definecolor{CTT}{HTML}{627380}
\definecolor{Gray}{HTML}{808080}

\addplot[only marks,color=black,fill=black,mark=+,mark size=3pt,line width=1.5pt]
    coordinates { (0,0)  };
    \addlegendentry{Ref.}

\addplot[only marks,color=black,fill=black,mark=*,mark size=3pt,line width=0.25pt]
    coordinates { (29.48,-37.41)  };
    \addlegendentry{CTP~\cite{Kraenzler21}} 
    
\addplot[only marks,color=black,fill=EA,mark=*,mark size=3pt,line width=0.25pt]
    coordinates { (28.62,-44.84)  };
    \addlegendentry{\ctp{EA}}

\addplot[only marks,color=black,fill=CA,mark=*,mark size=3pt,line width=0.25pt]
    coordinates { (22.19,-43.77)  };
    \addlegendentry{\ctp{CA}}

\addplot[only marks,color=black,fill=EB,mark=*,mark size=3pt,line width=0.25pt]
    coordinates { (27.00,-45.31)  };
    \addlegendentry{\ctp{E1}} 
    
\addplot[only marks,color=black,fill=CB,mark=*,mark size=3pt,line width=0.25pt]
    coordinates { (19.66,-41.34)  };
    \addlegendentry{\ctp{C1}}

\addplot[only marks,color=black,fill=white,mark=*,mark size=3pt,line width=0.25pt]
    coordinates { (-0.25,-4.86)  };
    \addlegendentry{\ctp{LBE 1-4}}

\addplot[only marks,color=black,fill=black,mark=*,mark size=3pt,line width=0.25pt]
    coordinates { (77.20,-41.68)  };

\addplot[only marks,color=black,mark=diamond*,mark size=5pt,line width=0.25pt]
    coordinates { (15.37,-28.28)  };

\addplot[only marks,color=black,fill=EA,mark=diamond*,mark size=5pt,line width=0.25pt]
    coordinates { (10.30,-40.37)  };

\addplot[only marks,color=black,fill=EB,mark=diamond*,mark size=5pt,line width=0.25pt]
    coordinates { (9.28,-37.65)  };

\addplot[only marks,color=black,fill=CA,mark=diamond*,mark size=5pt,line width=0.25pt]
    coordinates { (9.42,-38.51)  };

\addplot[only marks,color=black,fill=CB,mark=diamond*,mark size=5pt,line width=0.25pt]
    coordinates { (7.33,-30.19)  };

\addplot[only marks,color=black,fill=white,mark=*,mark size=3pt,line width=0.25pt]
    coordinates { (1.45,-11.41)  };

\addplot[only marks,color=black,fill=white,mark=*,mark size=3pt,line width=0.25pt]
    coordinates { (2.54,-17.55)  };

\addplot[only marks,color=black,fill=white,mark=*,mark size=3pt,line width=0.25pt]
    coordinates { (4.88,-25.54)  };

 \addplot [ 
   color=green,
    mark=none,
    mark size=2pt,
    line width=1pt,
    ] coordinates { (30,-45.31) (27.00,-45.31)   
	                (22.19,-43.77) (19.66,-41.34)
	                (10.30,-40.37) 
	                (9.42,-38.51) (9.28,-37.65)  
                    (7.33,-30.19) 
                    (4.88,-25.54) (2.54,-17.55)
                    (1.45,-11.41)                     
                    (-0.25,-4.86)  (-0.25,2)				
			  };
			  
\end{axis}
\end{tikzpicture}
\caption{ Evaluation of the proposed CTPs in terms of BDDE-VMAF (vertical axis) and BDR-VMAF (horizontal axis) for the JVET sequence set. For each of the four optimization strategies, we show a profile with the maximum energy efficiency (circle shaped marker) and a profile with optimal joint efficiency of energy and compression efficiency (diamond markers). Additionally, we show the four proposed low bit rate profiles and a Pareto front for the maximum energy saving at a given bit rate increase.}
\label{fig:JVETValidation}
\end{figure}
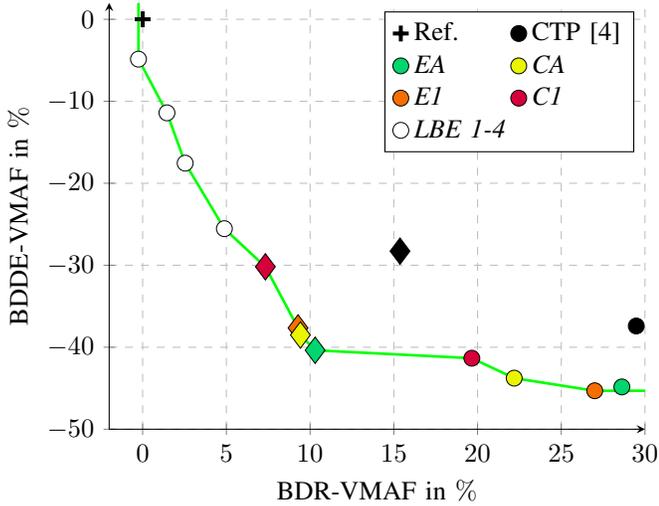
\vspace{-0.2cm}

\end{center}

In addition, we also propose four CTPs with a bit rate increase of less than $5\%$. These CTPs are enumerated in ascending order of the additional bit rate increase. For the \ctp{LBE 1} CTP, we measure an energy optimization of $\smi4.86\%$ and a bit rate optimization of $\smi0.25\%$. For the \ctp{LBE 4} CTP, the energy demand is reduced by over $\smi25\%$ with an additional bit rate that is comparable to the preset \ctp{slow}. 

For the UVG set, the observations are very similar to the JVET set. For the \ctp{E1 EE} CTP, we determine an improvement of $\smi47.12\%$ in terms of BDDE-VMAF. The \ctp{EBE} CTPs improve the energy demand by approximately $\smi37\%$ for the same additional bit rate as the \ctp{medium} preset. The \ctp{LBE 4} CTP has a similar compression efficiency as the preset \ctp{slow} and achieves an energy demand improvement of $\smi19.94\%$.

The results of BDDE-PSNR in Table~\ref{tab:Validation} are slightly higher than the corresponding BDDE-VMAF value. However, for BDR-PSNR, the differences are significantly larger in relation to BDR-VMAF. Especially, for \ctp{EBE} CTPs, that are also optimized on BDR-VMAF, a difference of up to $10\%$ can be observed. This indicates that the optimization strategies can optimize perceptual quality that is measured by VMAF.

In Table~\ref{tab:CTPUsage}, the usage of each coding tool is shown. For the \ctp{E1 EE} CTP, it can be observed that all in-loop filter coding tools and several others such as DMVR and BDOF are not used. However, for the \ctp{C1 EBE} profile, we determine that ALF, which is an in-loop filter, is used by the CTP along with other coding tools such as AFFINE, MIP, and LFNST. Those coding tools offer a reasonable trade off in terms of rate distortion and energy efficiency.

In summary, the joint optimization of the compression and energy efficiency provided CTPs with a low increase in bit rate and a significant improvement in energy efficiency. The usage of the \ctp{C1} optimization strategy is useful for low bit rate increases. The highest improvement in terms of energy efficiency is achieved with the \ctp{E1} strategy, but other strategies provided similar improvements in terms of energy efficiency.

	\begin{table}[!t]
 
       \caption{ Evaluation of the energy and compression efficiency of all proposed CTPs for the JVET (a) and UVG (b) set. For each CTP, we provide all four BD metrics that were used in this paper. The propsed CTPs are given in ascending order of the BDDE-VMAF value.}
       \vspace{-0.4cm}
\label{tab:Validation}
       \begin{center}
       \subfloat[JVET]{
       \begin{tabular}{l || c : c | c : c } 
         & \multicolumn{2}{c|}{VMAF} &  \multicolumn{2}{c}{PSNR}  \\
         & BDR & BDDE & BDR & BDDE    \\
         \hline 
         \ctp{Slow} & 5.31\,\% & -1.31\,\% & 6.03\,\% & -1.15\,\%\\
         \ctp{Medium} & 10.89\,\% & -0.97\,\% & 11.97\,\% & -0.78\,\%\\
          HEVC & 77.20\,\% & -41.68\,\% & 42.66\,\% & -43.20\,\%\\
         \hdashline
         \ctp{EE}~\cite{Kraenzler21} & 30.17\,\% & -37.66\,\% & 27.62\,\% & -36.69\,\%\\
         \ctp{EBE}~\cite{Kraenzler21} & 15.37\,\% & -28.28\,\% & 20.33\,\% & -26.79\,\%\\
         \hline
         \ctp{E1 EE} & 27.00\,\% & -45.31\,\% & 27.13\,\% & -44.50\,\%\\
         \ctp{EA EE} & 28.62\,\% & -44.84\,\% & 29.27\,\% & -44.01\,\%\\
         \ctp{CA EE} & 22.19\,\% & -43.77\,\% & 24.21\,\% & -42.66\,\%\\
         \ctp{C1 EE} & 19.66\,\% & -41.34\,\% & 18.19\,\% & -40.48\,\%\\
	\hdashline         
         \ctp{EA EBE} & 10.30\,\% & -40.37\,\% & 20.47\,\% & -38.29\,\%\\
         \ctp{CA EBE} & 9.42\,\% & -38.51\,\% & 18.85\,\% & -36.48\,\%\\
         \ctp{E1 EBE} & 9.28\,\% & -37.65\,\% & 15.83\,\% & -36.02\,\%\\
         \ctp{C1 EBE} & 7.33\,\% & -30.19\,\% & 11.37\,\% & -29.02\,\%\\
         \hdashline
         
         \ctp{LBE 4} & 4.88\,\% & -25.54\,\% & 10.06\,\% & -24.48\,\%\\
         \ctp{LBE 3} & 2.54\,\% & -17.55\,\% & 2.85\,\% & -17.30\,\%\\
         \ctp{LBE 2} & 1.45\,\% & -11.41\,\% & 3.59\,\% & -11.30\,\%\\
         \ctp{LBE 1} & -0.25\,\% & -4.86\,\% & -0.19\,\% & -4.81\,\%\\

       \end{tabular}     
       }
       \vspace{-0.1cm}
       \subfloat[UVG]{
          \begin{tabular}{ l || c : c | c : c } 
          & \multicolumn{2}{c|}{VMAF} &  \multicolumn{2}{c}{PSNR} \\
          & BDR & BDDE & BDR & BDDE  \\
         \hline
         \ctp{Slow} & 3.45\,\% & -0.68\,\% & 3.86\,\% & -0.94\,\%\\
         \ctp{Medium} & 7.88\,\% & -0.69\,\% & 8.44\,\% & -0.96\,\%\\
         HEVC & 85.23\,\% & -32.87\,\% & 41.63\,\% & -35.18\,\%\\
                  \hdashline
         \ctp{EE}~\cite{Kraenzler21} & 23.55\,\% & -39.37\,\% & 22.86\,\% & -38.38\,\%\\
         \ctp{EBE}~\cite{Kraenzler21} & 11.48\,\% & -27.50\,\% & 15.92\,\% & -26.89\,\%\\
         \hline
	    \ctp{E1 EE} & 21.29\,\% & -47.12\,\% & 23.12\,\% & -46.84\,\%\\
         \ctp{EA EE} & 20.62\,\% & -46.68\,\% & 23.81\,\% & -46.37\,\%\\
         \ctp{CA EE} & 15.62\,\% & -44.77\,\% & 20.28\,\% & -44.34\,\%\\
         \ctp{C1 EE} & 15.46\,\% & -44.14\,\% & 16.85\,\% & -43.68\,\%\\
         \hdashline      
         \ctp{EA EBE} & 7.87\,\% & -37.83\,\% & 17.65\,\% & -37.11\,\%\\
	    \ctp{CA EBE} & 8.19\,\% & -36.24\,\% & 19.51\,\% & -35.16\,\%\\
	    \ctp{E1 EBE} & 6.14\,\% & -35.15\,\% & 11.34\,\% & -34.76\,\%\\         
         \ctp{C1 EBE} & 6.10\,\% & -28.64\,\% & 10.14\,\% & -28.21\,\%\\
		\hdashline        
	    \ctp{LBE 4} & 3.74\,\% & -19.94\,\% & 9.63\,\% & -19.52\,\%\\         
         \ctp{LBE 3} & 2.60\,\% & -16.40\,\% & 1.61\,\% & -16.34\,\%\\
         \ctp{LBE 2} & 2.47\,\% & -11.78\,\% & 4.11\,\% & -11.71\,\%\\
         \ctp{LBE 1} & -0.12\,\% & -4.65\,\% & -0.11\,\% & -4.55\,\%\\
       \end{tabular} 
       }
       \end{center} 
     \end{table}

\begin{table*}[!t]
\def\arraystretch{1.05}
\caption{ Proposed EE and EBE profiles for all CTP as derived by the DSE algorithm. Furthermore, four \ctp{LBE} CTPs are given. Coding tools that are used in a CTP are denoted by \yes and disabled coding tools by \no.}
\label{tab:EnergyEfficientConfigs}
\begin{center}
\begin{tabular}{ c l | cc | cc | cc | cc | cc | cccc} 
     &       &  \multicolumn{2}{c|}{CTPs \cite{Kraenzler21}} &  \multicolumn{2}{c|}{\ctp{EA}} &  \multicolumn{2}{c|}{\ctp{E1}} &   \multicolumn{2}{c|}{\ctp{CA}} &  \multicolumn{2}{c|}{\ctp{C1}} & \multicolumn{4}{c}{\ctp{LBE}} \\
 & Tool      &  \ctp{EE} & \ctp{EBE} & \ctp{EE} & \ctp{EBE} &  \ctp{EE} & \ctp{EBE} &  \ctp{EE} & \ctp{EBE} & \ctp{EE} & \ctp{EBE} & \ctp{LBE 1} & \ctp{LBE 2} & \ctp{LBE 3} & \ctp{LBE 4} \\
\hline\hline
      \multirow{4}{*}{ Intra} & CCLM & \no  & \no  & \no  & \no  & \no  & \yes  & \no  & \no  & \no  & \no  & \yes  & \yes  & \yes  & \no  \\ 
      & ISP & \no  & \no  & \no  & \no  & \no  & \yes  & \no  & \no  & \yes  & \yes  & \yes  & \yes  & \yes  & \yes  \\ 
      & MIP & \no  & \no  & \no  & \no  & \no  & \yes  & \no  & \no  & \yes  & \yes  & \yes  & \yes  & \yes  & \yes  \\ 
      & MRL & \yes  & \yes  & \no  & \yes  & \yes  & \yes  & \yes  & \yes  & \yes  & \yes  & \yes  & \yes  & \yes  & \yes  \\ 
\hline \hline
  \multirow{11}{*}{Inter} & AFFINE & \no  & \no  & \no  & \no  & \no  & \no  & \no  & \yes  & \yes  & \yes  & \yes  & \yes  & \yes  & \yes  \\ 
      & AMVR & \yes  & \yes  & \yes  & \yes  & \yes  & \yes  & \yes  & \yes  & \yes  & \yes  & \yes  & \yes  & \yes  & \yes  \\ 
      & BCW & \no  & \yes  & \no  & \yes  & \yes  & \yes  & \yes  & \no  & \yes  & \yes  & \yes  & \yes  & \yes  & \yes  \\ 
      & BDOF& \no  & \no  & \no  & \yes  & \no  & \yes  & \no  & \yes  & \no  & \no  & \yes  & \yes  & \yes  & \yes  \\ 
      & CIIP & \no  & \no  & \no  & \no  & \no  & \no  & \no  & \no  & \no  & \no  & \yes  & \yes  & \yes  & \no  \\ 
      & DMVR & \no  & \no  & \no  & \no  & \no  & \no  & \no  & \no  & \no  & \no  & \yes  & \yes  & \yes  & \no  \\ 
      & GPM & \yes  & \yes  & \yes  & \yes  & \yes  & \yes  & \yes  & \yes  & \yes  & \yes  & \yes  & \yes  & \yes  & \yes  \\ 
      & MMVD & \yes  & \yes  & \no  & \yes  & \yes  & \yes  & \yes  & \yes  & \yes  & \yes  & \yes  & \yes  & \yes  & \yes  \\ 
      & PROF & \no  & \no  & \yes  & \no  & \yes  & \yes  & \no  & \no  & \no  & \no  & \yes  & \yes  & \yes  & \no  \\ 
      & SBTMVP & \no  & \no  & \no  & \yes  & \no  & \no  & \yes  & \yes  & \yes  & \yes  & \yes  & \yes  & \yes  & \yes  \\ 
      & SMVD & \no  & \no  & \yes  & \yes  & \yes  & \yes  & \no  & \no  & \yes  & \yes  & \yes  & \yes  & \yes  & \yes  \\ 
       \hline \hline     
    & DQ & \yes  & \yes  & \yes  & \yes  & \yes  & \yes  & \yes  & \yes  & \yes  & \yes  & \yes  & \yes  & \yes  & \yes  \\ 
      & JCCR & \yes  & \yes  & \yes  & \yes  & \yes  & \yes  & \yes  & \yes  & \yes  & \yes  & \yes  & \yes  & \yes  & \yes  \\ 
   Transform \&  & LFNST & \no  & \no  & \no  & \yes  & \no  & \yes  & \yes  & \yes  & \yes  & \yes  & \yes  & \yes  & \yes  & \yes  \\ 
   Quantization    & MTS & \no  & \no  & \no  & \yes  & \no  & \yes  & \yes  & \yes  & \yes  & \yes  & \yes  & \yes  & \yes  & \yes  \\ 
     & SBT & \no  & \no  & \yes  & \no  & \yes  & \yes  & \yes  & \yes  & \yes  & \yes  & \yes  & \yes  & \yes  & \yes  \\ 
      & TSRC & \yes  & \yes  & \yes  & \no  & \yes  & \yes  & \no  & \no  & \yes  & \yes  & \yes  & \yes  & \yes  & \yes  \\ 
      \hline \hline
      \multirow{5}{*}{In-Loop Filter}  & ALF & \no  & \yes  & \no  & \no  & \no  & \no  & \no  & \no  & \no  & \yes  & \yes  & \yes  & \yes  & \yes  \\ 
      & CCALF & \no  & \no  & \yes  & \yes  & \no  & \yes  & \no  & \no  & \no  & \no  & \yes  & \no  & \yes  & \no  \\ 
      & DBF & \no  & \no  & \no  & \no  & \no  & \no  & \no  & \no  & \no  & \no  & \yes  & \no  & \no  & \no  \\ 
      & LMCS & \no  & \no  & \no  & \no  & \no  & \no  & \no  & \no  & \no  & \no  & \yes  & \yes  & \no  & \no  \\ 
      & SAO & \yes  & \yes  & \no  & \no  & \no  & \no  & \no  & \no  & \no  & \no  & \yes  & \yes  & \yes  & \no  \\ 
  \hline \hline
   \multirow{4}{*}{Others} & BDPCM    & \yes  & \yes  & \no  & \no  & \yes  & \yes  & \no  & \no  & \yes  & \no  & \yes  & \yes  & \yes  & \yes  \\ 
      & IBC & \yes  & \yes  & \no  & \no  & \no  & \no  & \no  & \no  & \no  & \no  & \no  & \no  & \no  & \no  \\ 
      & CST & \yes  & \yes  & \yes  & \yes  & \yes  & \yes  & \no  & \no  & \yes  & \yes  & \yes  & \yes  & \yes  & \yes  \\ 
      & MCTF & \yes  & \yes  & \yes  & \yes  & \yes  & \yes  & \yes  & \yes  & \yes  & \yes  & \yes  & \yes  & \yes  & \yes  \\ 
         \end{tabular}
\end{center}
\label{tab:CTPUsage}
\vspace{-0.3cm}
\end{table*}

\section{Conclusion}

\label{sec:conclusion}
In this paper, we optimized the DSE algorithm by applying several optimization strategies. It was shown that the \ctp{E1} strategy has the highest improvement in energy efficiency. Furthermore, we showed that the joint optimization of the energy and compression efficiency provides energy savings close to $\smi40\%$ with an additional bit rate of 10\%. For a low bit rate increase of less than 10\%, the strategies provide a plethora of coding tool profiles that can be selected. In summary, we proposed 16 coding tool profiles that optimize the energy demand by $\smi25\%$ for 5\% additional bit rate, by $\smi40\%$ for 10\% additional bit rate, and by up to $\smi45\%$ with 27\% additional bit rate for the JVET sequence set. 

In future work, we will evaluate the DSE algorithm on the encoder complexity optimization. Furthermore, we will study the possibilities of the DSE to reduce the decoder complexity of other video codecs. Finally, a method shall be developed that comprises all optimization strategies to target certain pre-defined bit rate increases.

\bibliographystyle{IEEEtran}

\end{document}